# Quantum Interference Effects in InAs Semiconductor Nanowires

Yong-Joo Doh,[1,*] Aarnoud L. Roest,[2] Erik P. A. M. Bakkers,[2] Silvano De Franceschi,[3,4] and Leo P. Kouwenhoven[1]

[1] Kavli Institute of Nanoscience, Delft University of Technology, PO Box 5046, 2600 GA Delft, Nehterlands

[2] Philips Research Laboratories, Professor Holstlaan 4, 5656 AA Eindhoven, Netherlands

[3] LaTEQs laboratory, DSM/DRFMC/SPSMS, CEA-Grenoble, 17 rue des Martyrs, 38054 Grenoble, France

[4] TASC laboratory, CNR-INFM, S.S. 14, Km 163.5, 34012 Trieste, Italy

[*] Present address: National CRI Center for Semiconductor Nanorods, Department of Materials Science and Engineering, Pohang University of Science and Technology, Republic of Korea (e-mail: yjdoh@postech.ac.kr)

## Abstract

We report quantum interference effects in InAs semiconductor nanowires strongly coupled to superconducting electrodes. In the normal state, universal conductance fluctuations are investigated as a function of magnetic field, temperature, bias and gate voltage. The results are found to be in good agreement with theoretical predictions for weakly disordered one-dimensional conductors. In the superconducting state, the fluctuation amplitude is enhanced by a factor up to ~ 1.6, which is attributed to a doubling of charge transport via Andreev reflection. At a temperature of 4.2 K, well above the Thouless temperature, conductance fluctuations are almost entirely suppressed, and the nanowire conductance exhibits anomalous quantization in steps of $e^2/h$.

Chemically grown semiconductor nanowires can provide a mesoscopic system to study quantum confinement and interference effects at low temperature, which is a promising platform to develop novel quantum devices. In Coulomb blockade regime, single-electron tunneling devices [1, 2] and few-electron quantum dots [3] have been realized successfully from various nanowires. In strong coupling regime, Kondo effect [4], weak localization [5] and universal conductance fluctuations [4-6] have been observed using InAs nanowires. With highly transparent contacts to conventional superconductors, a supercurrent can flow through the semiconductor nanowire to enable Josephson field-effect transistors [6, 7] and gate-tunable superconducting quantum interference devices [8] at the nanoscale.

Universal conductance fluctuations (UCF) are caused by the quantum interference of multiply scattered electronic wavefunctions in a weakly disordered conductor, giving rise to aperiodic conductance fluctuations as a function of magnetic field and Fermi energy [9, 10]. As the sample size, $L$, becomes smaller than the phase coherence length $L_\phi = (D\tau_\phi)^{1/2}$, where $D$ is the electron diffusion constant and $\tau_\phi$ the inelastic scattering time, the root-mean-square (rms) amplitude of the fluctuations is of order $e^2/h$ independent of the degree of disorder [9, 10]. Here, $e$ is the electric charge and $h$ Planck's constant. When the mesoscopic normal conductor is brought into contact with a superconductor, the phase-coherent electronic transport is expected to incorporate superconducting correlations [11], resulting in a combination of UCF and Andreev reflection [12]. Following some initial pioneering work [13] based on all-metallic systems, further investigation of such a fundamental phenomenon has remained an experimental challenge [14].

In this Letter, we investigate the conductance fluctuations of InAs nanowires, contacted with superconducting Al electrodes, as a function of magnetic field $B$, back-gate voltage $V_g$, bias $V$, and temperature $T$. The magnetoconductance data show reproducible and aperiodic fluctuations with a characteristic amplitude of order $e^2/h$. We estimate the phase-coherence length to be $L_\phi \sim 100$ nm at $T = 30$ mK. The autocorrelation function of the magnetoconductance data decays on a field scale consistent with this value of $L_\phi$, In the superconducting state, the amplitude of conductance fluctuations is enhanced by a factor up to ~ 1.6 at low bias below the superconducting energy gap $2\Delta/e$, which is attributed to a participation of Andreev reflected holes in UCF. As we increase temperature above the Thouless temperature $E_{Th}/k_B \sim 1.2$ K, where $k_B$ is the Boltzmann constant, the conductance fluctuations are suppressed by thermal dephasing. With UCF almost washed out at $T = 4.2$ K, anomalous conductance plateaus at multiples of $e^2/h$ are observed as a function of $V_g$, which are insensitive to the application of a perpendicular magnetic field up to $B \sim 2$ T. The possible origin of this anomalous quantization is discussed.

Single crystalline InAs nanowires are grown via laser-assisted vapor-liquid-solid method. After depositing the nanowires on a degenerately doped p-type silicon substrate with a 250-nm-thick surface oxide, superconducting contacts are formed using Ti(10 nm)/Al(120 nm). Details on the nanowire growth and device fabrication have been published elsewhere [6]. The linear conductance $G$ of the nanowire device, corresponding to the inverse of a dynamic resistance $(dV/dI)^{-1}$, is measured using an AC lock-in technique, in which the AC voltage across the sample is kept below $k_BT/e$ to avoid electron heating. To reduce the external noise effects, the measurement leads were filtered by $\Pi$ filters at room temperature and low-pass RC and copper powder filters at the temperature of the mixing chamber in a dilution refrigerator.

Typical magnetoconductance data at $T = 30$ mK are shown in Fig. 1a. The magnetic field, $B$, was applied parallel (perpendicular) to the nanowire axis for device **D1** (**D2**). The overshoot of $G(B)$ curve near $B = 0$ T is due to a supercurrent induced by the superconducting proximity effect. Regardless of the field direction, reproducible and aperiodic conductance fluctuations are observed as a function of magnetic field. The peak-to-peak variation of magnetoconductance is about $e^2/h$, consistent with theoretical predictions [9, 10]. $G(B)$ curve is symmetric upon field reversal as expected for a mesoscopic two-probe measurement [15]. The rms magnitude of the magnetoconductance fluctuations is defined as $\mathrm{rms}(G_B) = <(G(B)-<G(B)>)^2>^{1/2}$, where the angular brackets refer to an average over magnetic field, resulting in $\mathrm{rms}(G_B) = 0.29$ (0.30) $e^2/h$ for **D1** (**D2**) with perpendicular magnetic field. In this average we disregarded the low-field data range ($|B| <$ 0.5 T) where the magnetoconductance is affected by weak localization/antilocalization [5] and superconducting proximity effect [6].

The phase coherence length $L_\phi$ can be obtained from the analysis of the autocorrelation function of $G(B)$, which is defined as $F(\Delta B) = <G(B)G(B+\Delta B)> - <G(B)>^2$ with $\Delta B$ a lag parameter in magnetic field [9, 10]. $F(\Delta B)$ is expected to have a peak at $\Delta B = 0$. The half-width at half height of this peak corresponds to a magnetic correlation length, $B_c$, over which the phases of interference paths become uncorrelated with those at the initial field. Figure 1b shows the positive side ($\Delta B > 0$) of the autocorrelation function. From the data obtained in perpendicular magnetic field (open dots) we find $B_c = 0.21$ T and 0.18 T for device D1 and D2, respectively. According to theoretical calculations for a quasi-one-dimensional conductor [11], the correlation field is expected to be inversely proportional to the coherence length, i.e. $B_c = 0.42 \, \Phi_0/(wL_\phi)$, where $\Phi_0 = h/e$ is the one-electron flux quantum and $w$ a width corresponding to the nanowire diameter (80 nm). From $B_c \approx 0.2$ T we find $L_\phi \approx 100$ nm. This value is smaller than the one obtained from weak localization/antilocalization measurements in similar nanowires [5]. Since the Fermi wave number $k_F$ is estimated to be $\sim 5 \times 10^6$ cm$^{-1}$ from the carrier concentration $n_s \sim 6 \times 10^{18}$ cm$^{-3}$ [6], our assumption of the quasi-one-dimensional conductor is satisfied with $k_F l \gg 1$ where $l = 10 - 100$ nm is an elastic mean free path [5, 6]. Similar $F(\Delta B)$ curves are obtained with the magnetic field applied parallel to the nanowire axis, in contrast with previous results for multi-walled carbon nanotubes [15]. We argue that the seemingly weak dependence of $B_c$ on the field direction reflects the fact that $L_\phi$ is very close to the nanowire diameter.

The obtained values of $L_\phi$ can be used to verify the consistency between the observed UCF amplitude and the corresponding theoretical expectation. When the coherence length $L_\phi$ is much shorter than the sample size $L$, the nanowire can be considered as a series of uncorrelated segments of length $L_\phi$. The fluctuations are described by rms$(G_B) = 2.45(L_\phi/L)^{3/2}$ [16], resulting in $0.27e^2/h$ for **D1**, and 2.1 $e^2/h$ for **D2**. While the first value is in good agreement with the measured UCF amplitude, in the second case we find a significant discrepancy which we interpret as the result of an effective channel length substantially larger than the lithographic distance between the contacts. To support this interpretation, we note that contact electrodes are 500 nm wide and, within the same contact, the transparency of the metal-nanowire interface can be strongly inhomogeneous. The hypothesis of a larger channel length for **D2** is further substantiated by the relatively small value of the conductance ($\sim 36e^2/h$) as compared to **D1** ($\sim 22.5e^2/h$).

In the absence of magnetic field the conductance fluctuations are observed as a function of $V_g$, since a change of chemical potential induced by $V_g$ is equivalent to a change in impurity configuration in the nanowire [17]. Figure 2a shows bias-dependent $\delta G(V_g)$ curves, in which the background conductance was subtracted from the raw data of $G(V_g)$ after a second order polynomial fit. With increasing bias $V$, the fluctuation amplitude decreases substantially while its pattern is deformed progressively from the initial one at the low bias. The bias-dependent rms amplitude of the fluctuations, $\text{rms}(G_g) = \langle(\delta G(V_g))^2\rangle^{1/2}$, with angular brackets referring to an average over $V_g$, is depicted in Fig. 2b. The rms amplitude drops abruptly as the bias voltage exceeds the superconducting energy gap of the electrodes, $V_{\text{gap}} = 2\Delta/e \sim 0.23$ mV. Above $V_{\text{gap}}$, however, $\text{rms}(G_g)$ remains larger than the $\text{rms}(G_B)$ extracted from Fig. 1a. We ascribe this residual enhancement to which is mostly attributed to the presence of time-reversal symmetry at zero magnetic field. The enhancement factor in the normal state, $\text{rms}(G_g)|_{V=0.44\text{mV}}/\text{rms}(G_B)$, is about 1.7 for **D1**, which is quite close to the theoretical expectation of 1.41 [11].

We suggest that another enhancement of $\text{rms}(G_g)$ value at low bias below $V_{\text{gap}}$ is a direct evidence of the interplay between UCF and Andreev reflection. The inset of Fig. 2b shows a typical dynamic conductance $dI/dV(V)$ curve at low temperature far below the superconducting transition temperature $T_c = 1.1$ K of Al electrode. The overall conductance enhancement at low bias below $V_{\text{gap}}$ is caused by the Andreev reflection at the interface between the InAs nanowire and the superconducting electrodes, where the incident normal electron is retro-reflected as a phase conjugated hole [12]. Multiple conductance peaks at $V_m = V_{\text{gap}}/m$ with $m = 1, 2, 3$ occur when the Andreev-reflected hole is reflected again as a normal electron at the opposite interface, or vice versa [18]. Finally additional elementary charges of Andreev-reflected holes are driven into a weakly disordered system of InAs nanowire to increase the rms amplitude of conductance fluctuations at low bias below $V_{\text{gap}}$. For a phase-coherent segment of $L_\phi$ near the interface, the enhancement factor $\alpha$ of the rms amplitude of UCF in the superconducting state relative to the one in the normal state is given to be about $\alpha = 2.08$ in theory with the assumption of time reversal symmetry [11]. Thus, for the whole nanowire segment of $L$ between two superconducting contacts, the total enhancement factor $\gamma$ is obtained to be $(1+2(\alpha^2-1)/N_\phi)^{1/2}$ with $N_\phi = L/L_\phi$, giving rise to $\gamma = 1.59$ for **D1**, which is quite close to the experimental value of $\text{rms}(G_g)|_{V=0.1\text{mV}}/\text{rms}(G_g)|_{V=0.44\text{mV}} = 1.57$ in Fig. 2b.

Another characteristic length scale determining coherent electronic transport is the thermal length, defined as $L_T = (hD/2\pi k_B T)^{1/2}$. Using $D = 80$ cm$^2$/s [6], we obtain $L_T = 1.3$ μm at $T = 30$ mK. Since this is much longer than $L$ and $L_\phi$, we have so far ignored thermal smearing as a dephasing mechanism. We now discuss the effect of temperature. To investigate the *T*-dependence of conductance fluctuations in the absence of superconductivity we have applied a perpendicular magnetic field of $B = 0.1$ T, corresponding to a magnetic flux of $(0.2 - 0.9)\, \Phi_0$ in the nanowire segment. The ac bias for the lock-in measurement was kept below 10 μV in order to minimize electron heating effect. The results are shown in Fig. 3a. As temperature increases, rms($G_g$) is almost constant up to a critical temperature $T^* = 1.2$ K, above which it decreases substantially. Highly reproducible fluctuations of $\delta G(V_g)$ for $T < T^*$ are displayed in the inset of Fig. 3a over a large $V_g$ range. The critical temperature $T^*$ is linked to the Thouless energy, a characteristic energy scale $E_c$ for diffusive transport, which is defined as $E_c = hD/2\pi L^2$ [19]. For **D4**, $E_c$ is found to be 0.14 meV, compared to $k_B T^* = 0.10$ meV.

As the fluctuations have almost disappeared at $T = 4.2$ K, conductance plateaus emerges clearly at $G = me^2/h$ with $m = 3, 4, 5$ in the $G(V_g)$ curve, as shown in Fig. 3b. The conductance steps remain almost unaltered even after the application of a perpendicular magnetic field up to $B = 2.33$ T. The conductance values in the zero-field $G(V_g)$ curves of Fig. 3a are displayed as an intensity plot in Fig. 3c thereby emphasizing the anomalous conductance quantization in units of $e^2/h$. Here it should be noted that the linear conductance was measured in a four-terminal configuration to avoid any non-linear effects from the contact resistance. Two-terminal measurements for the nanowire segment of (B-C) and (C-D), however, show similar conductance plateaus after subtracting a contact resistance of 100 Ω, as shown in the upper inset of Fig. 3b for the segment of (C-D).

There are two distinctive features in our measurement differing from the quantized conductance of quantum point contacts in two-dimensional electron gas [20]. Firstly, we utilized only a back gate for the electrostatic depletion. It is speculated that an arbitrary quantum point contact is formed in the middle of the nanowire segment due to a nonuniform distribution of the electrostatic potential with the application of $V_g$ [21]. Secondly, the unit of conductance steps is $e^2/h$ rather than $2e^2/h$, where the factor of 2 corresponds to the spin degeneracy of the one-dimensional subbands [20]. It should be noted that at $B = 2.33$ T the Zeeman splitting $|g\mu_B B| = 2.02$ meV is larger than the thermal energy broadening $3.5k_\text{B}T = 1.26$ meV at $T = 4.2$ K (based on previous experiments [8] we have taken $g \approx -15$, the Lande $g$-factor in bulk InAs, while $\mu_\text{B}$ is the Bohr magneton). Similar anomalies in the conductance quantization have been reported for other one-dimensional nanostructures, such as carbon nanotubes [22], Ge/Si nanowires [23] and GaAs quantum wires [24]. The origin of the apparent lack of spin degeneracy is currently not understood. It has been proposed that a spontaneous spin polarization may occur in a one-dimensional electron gas at zero magnetic field [25]. More in-depth studies are necessary to shed light on this open issue.

In summary, we have investigated quantum interference effects in InAs semiconductor nanowires connecting superconducting metal contacts. In the normal state, conductance fluctuations as a function of magnetic field or gate voltage are in good agreement with theoretical predictions for a weakly disordered one-dimensional conductor. In the superconducting state, we have presented strong evidence of the interplay between UCF and the phase-coherent Andreev reflection phenomenon. Finally, following the temperature-induced suppression of UCF we have observed an anomalous conductance quantization whose physical origin remains to be clarified.

**Acknowledgement**
We gratefully acknowledge helpful discussions with C. W. J. Beenakker and Y. V. Nazarov. We acknowledge financial support from the EU through the HYSWITCH project and from the EUROCORES FoNE programme.

**References**

[1] S. De Franceschi *et al*., Appl. Phys. Lett. **83**, 344 (2003).

[2] C. Thelander *et al*., Appl. Phys. Lett. **83**, 2052 (2003).

[3] M. T. Björk *et al*., Nano Lett. **4**, 1621 (2004).

[4] T. S. Jespersen *et al*., Phys. Rev. B **74**, 233304 (2006).

[5] A. E. Hansen *et al*., Phys. Rev. B **71**, 205328 (2005).

[6] Y.-J. Doh *et al*., Science **309**, 272 (2005).

[7] J. Xiang *et al*., Nature Nanotechnology **1**, 208 (2006).

[8] J. A. van Dam *et al*., Nature (London) **442**, 667 (2006).

[9] P. A. Lee and A. D. Stone, Phys. Rev. Lett. **55**, 1622 (1985).

[10] B. L. Al'tshuler, Sov. Phys. JETP Lett. **12**, 648 (1985).

[11] C. W. J. Beenakker, Rev. Mod. Phys. **69**, 731 (1997).

[12] A. F. Andreev, Sov. Phys. JETP **19**, 1228 (1964).

[13] K. Hecker, H. Hegger, A. Altland, and K. Fiegle, Phys. Rev. Lett. **79**, 1547 (1997).

[14] S. G. den Hartog and B. J. van Wees, Phys. Rev. Lett. **80**, 5023 (1998).

[15] C. Schönenberger *et al*., Appl. Phys. A **69**, 283 (1999).

[16] C. W. J. Beenakker and H. van Houten, Solid State Phys. **44**, 1 (1991).

[17] H. T. Man and A. F. Morpurgo, Phys. Rev. Lett. **95**, 26801 (2005).

[18] M. Octavio, M. Tinkham, G. E. Blonder and T. M. Klapwijk, Phys. Rev. B **27**, 6739 (1983).

[19] D. J. Thouless, Phys. Rev. Lett. **39**, 1167 (1977).

[20] B. J. van Wees *et al*., Phys. Rev. Lett. **60**, 848 (1988).

[21] S. Heinze *et al*., Phys. Rev. Lett. **89**, 106801 (2002).

[22] M. J. Biercuk *et al*., Phys. Rev. Lett. **94**, 26801 (2005).

[23] W. Lu *et al*., Proc. Natl. Acad. Sci. U.S.A. **102**, 10046 (2005).

[24] R. Crook *et al*., Science **312**, 1359 (2006).

[25] K. J. Thomas *et al*., Phys. Rev. Lett. **77**, 135 (1996).

**Figure Captions**

**Figure 1**. (a) Linear conductance $G$ with magnetic field $B$ applied parallel (lower line for device **D1**) and perpendicular (upper for device **D2**) to the nanowire axis with $V_g = 0$ V at $T = 30$ mK. The overshoot of $G(B)$ curve below $B = 0.05$ T is due to a supercurrent through the nanowire. The source-drain spacing is $L = 440$ (**D1**) and 107 (**D2**) nm, respectively. (b) Autocorrelation functions $F(\Delta B)$, extracted from $G(B)$ curves at $V_g = 0$ V, for **D1** (red) and **D2** (black) with perpendicular (line) and parallel (circle) $B$. Inset: a scanning electron microscopy (SEM) picture of a typical device. The scale bar defines 1 μm.

**Figure 2.** (a) $\delta G(V_g)$ curve for **D1** with different bias $V = 0.11, 0.20, 0.44, 0.65, 0.88$ mV from bottom to top at $T = 22$ mK. Background conductance was subtracted from $G(V_g)$ data. Each graph is shifted for clarity. (b) Bias-dependent rms($G_g$) for device **D1** (circle) and **D3** (square). Inset: normalized $dI/dV(V)$ curve for **D2** at $T = 22$ mK. A series of conductance peaks at $V_m = V_{gap}/m$, where $V_{gap}$ is superconducting gap energy of Al electrode and $m = 1, 2, 3$, is caused by multiple Andreev reflection. Conductance overshoot near zero-bias indicates an existence of the supercurrent through the nanowire.

**Figure 3.** (a) Log-log plot of temperature dependence of rms($G_g$) for device **D1** (circle), **D2** (triangle), and **D4** (rectangle) with perpendicular magnetic field $B = 0.1$ T. Inset: temperature dependence of $\delta G(V_g)$ curve from **D1** with $T = 26$ (blue), 500 (green), and 1220 (red line) mK. (b) $G(V_g)$ curves from **D4** with increasing perpendicular magnetic field $B$ up to 2.33 T in increments of 0.137 T at $T = 4.2$ K. For clarity each plot is shifted by $V_g = + 2$ V. The conductance was measured in a four-terminal configuration. Lower inset: schematic view for the four-terminal measurement configuration. Current is injected at electrode A and removed by electrode D, while the voltage difference is measured between electrodes B and C. The electrode width is 500 nm and the channel length is $L = 120$ (A-B), 220 (B-C), 290 (C-D) nm, respectively. Upper inset: $G(V_g)$ curve at $T = 4.2$ K with $B = 0$ T for the nanowire segment of (C-D). (c) Intensity plot corresponding to the whole $G(V_g)$ curves in (b).

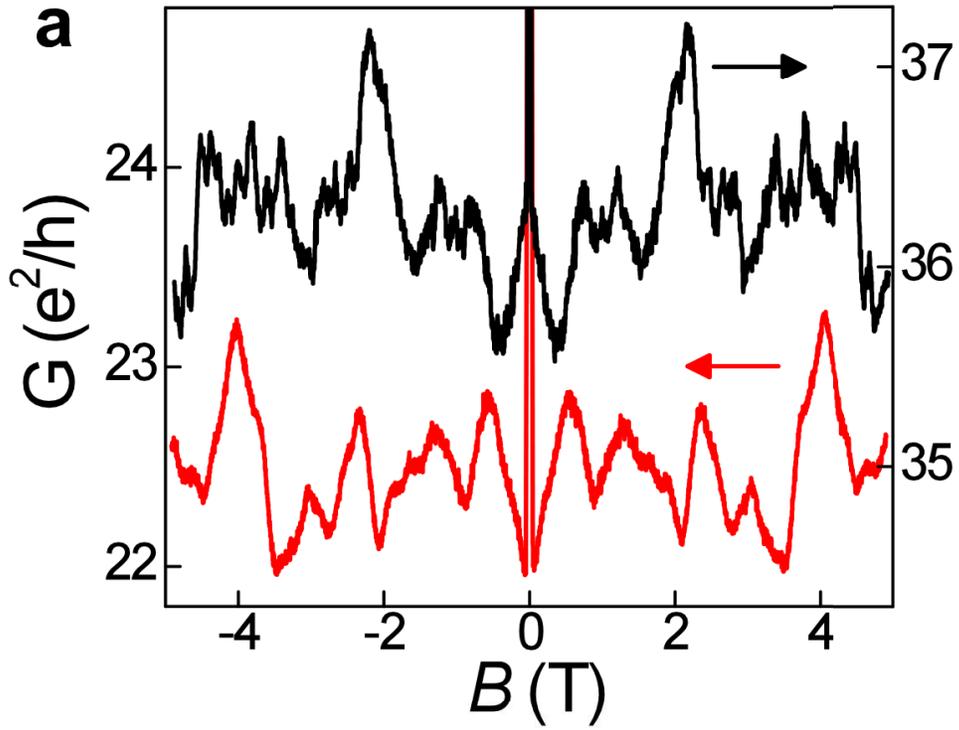

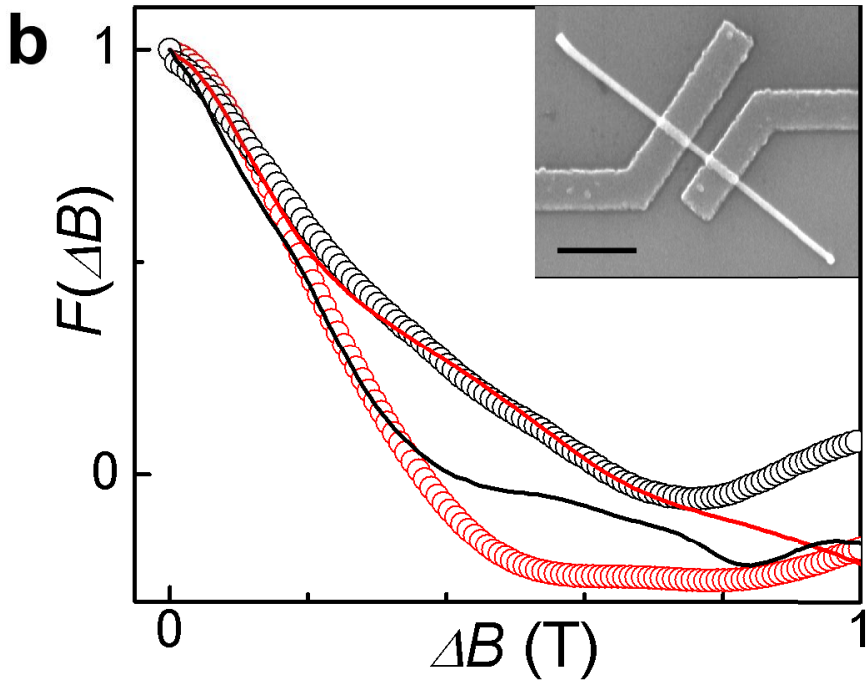

Doh et al., Fig. 2

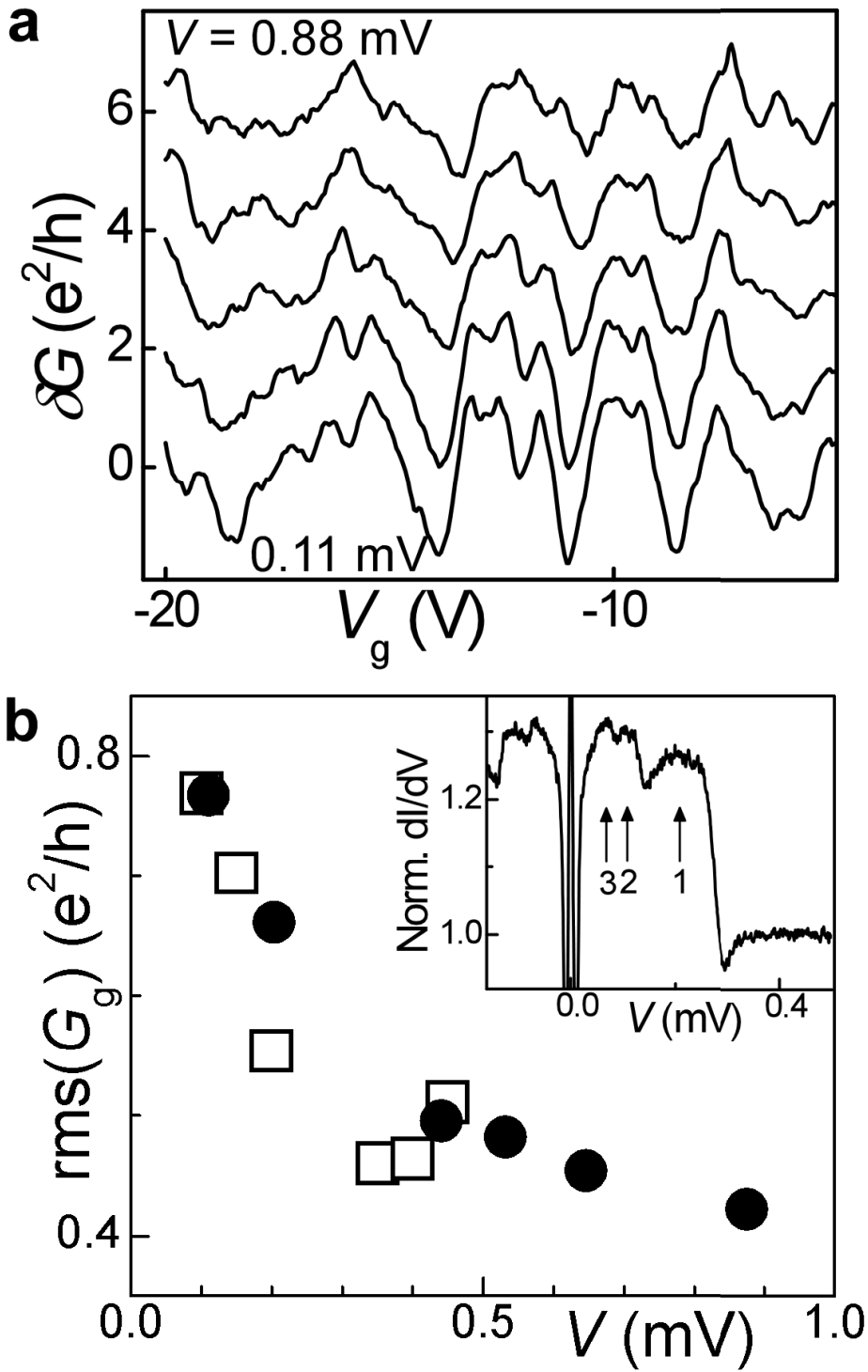

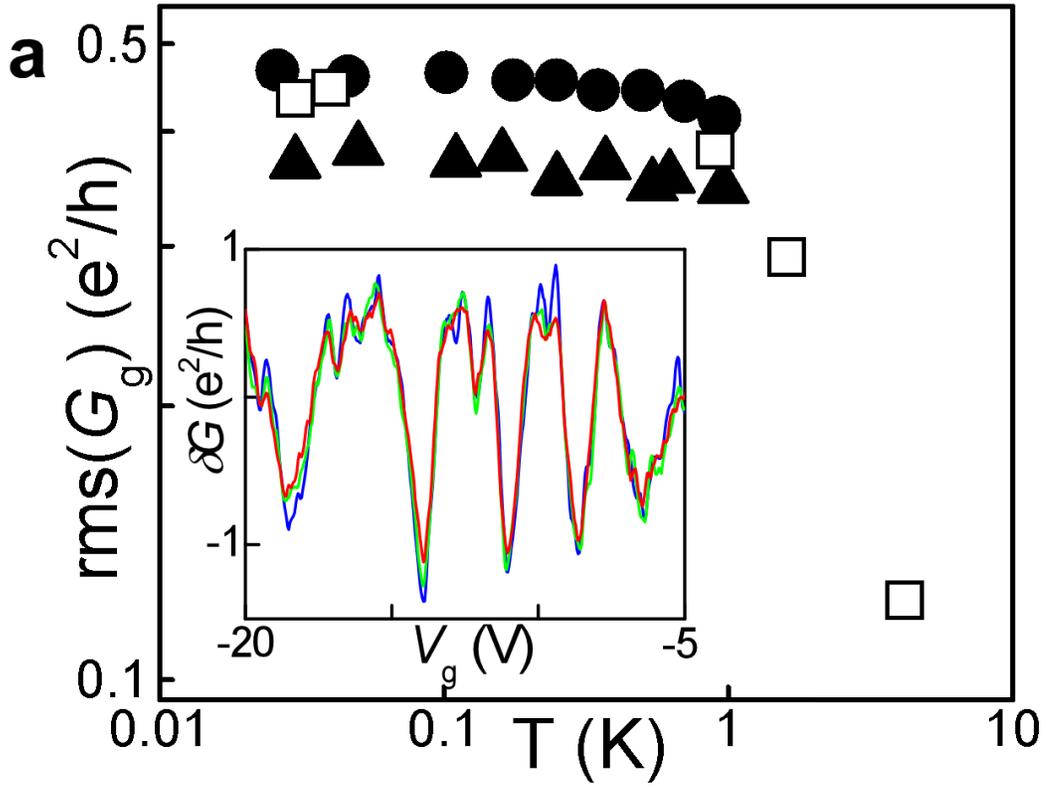
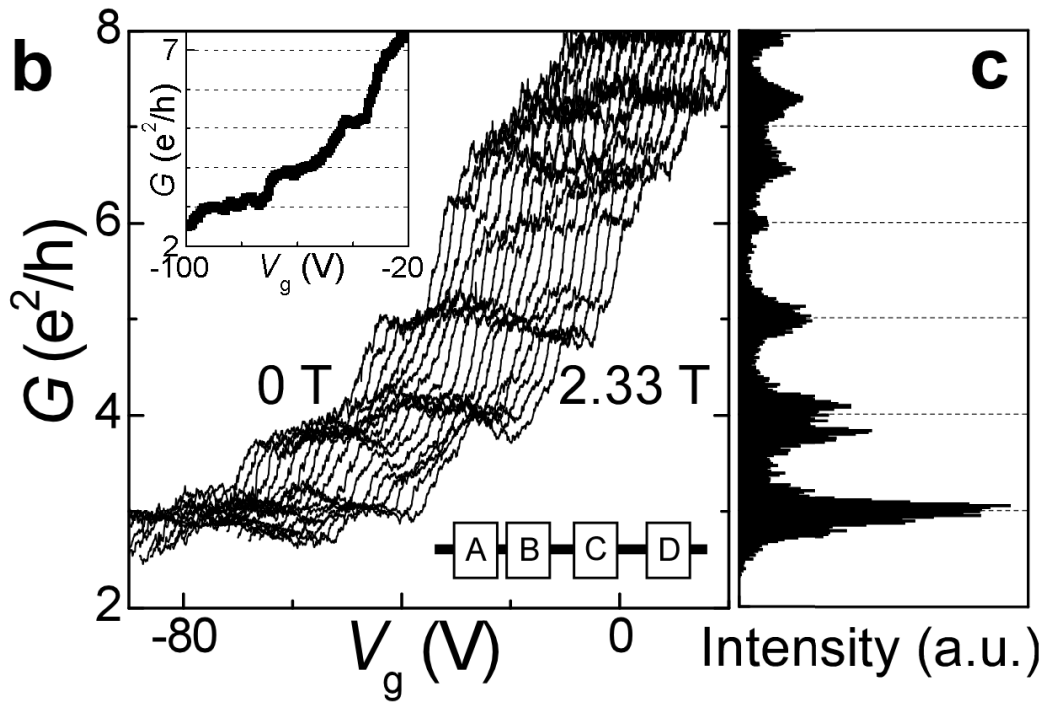